\documentclass[aps,prl,twocolumn,superscriptaddress]{revtex4-2}
\usepackage{amssymb}
\usepackage{amsmath}
\usepackage{graphicx}
\usepackage{array}
\usepackage{setspace}
\usepackage[colorlinks,linkcolor=blue,anchorcolor=blue,citecolor=red]{hyperref}
\usepackage[normalem]{ulem}

\begin{document}
\title{Microscopic Origin of Spin Splitting in Altermagnetic CrSb Thin Films}
\author{Mingyang Dai}
\affiliation{Institute for Applied Physics, University of Science and Technology Beijing, Beijing, 100083, China}
\author{Hongquan Song}
\affiliation{College of Physics and Telecommunication Engineering, Zhoukou Normal University, Zhoukou, 466001, China}
\author{Zhuo Kang}
\affiliation{Academy for Advanced Interdisciplinary Science and Technology, Engineering, Beijing Key Laboratory for Advanced Energy Materials and Technologies, University of Science and Technology Beijing, 100083, China}
\affiliation{State Key Laboratory for Advanced Metals and Materials, University of Science and Technology Beijing, Beijing, 100083, China}
\author{Yuanji Xu}
\email{yuanjixu@ustb.edu.cn}
\affiliation{Institute for Applied Physics, University of Science and Technology Beijing, Beijing, 100083, China}
\author{Fuyang Tian}
\email{fuyang@ustb.edu.cn}
\affiliation{Institute for Applied Physics, University of Science and Technology Beijing, Beijing, 100083, China}
\affiliation{State Key Laboratory for Advanced Metals and Materials, University of Science and Technology Beijing, Beijing, 100083, China}
\date{\today}

\begin{abstract}
Altermagnets have recently emerged as promising materials for spintronic applications owing to their momentum-dependent spin splitting. Among them, metallic CrSb is particularly attractive owing to its giant spin splitting and high N\'eel temperature. However, the microscopic origin of the distinct spin-splitting behaviors in bulk and thin-film CrSb remains unresolved. Here, we systematically investigate the electronic structures of CrSb slabs with different surface orientations using first-principles calculations. Although all considered slabs preserve spin-group symmetries compatible with altermagnetism, they exhibit markedly different electronic structures: the (2$\bar{1}\bar{1}$0) slab retains pronounced altermagnetic spin splitting, whereas the (0001) and (10$\bar{1}$0) slabs display nearly spin-degenerate bands. We demonstrate that dimensional reduction fundamentally changes the microscopic origin of altermagnetic spin splitting. Unlike bulk CrSb, altermagnetic spin splitting in thin films requires long-range inter-unit-cell coplanar Cr--Sb hopping, while Sb--Sb hopping provides an additional contribution. The preservation or suppression of these hopping pathways explains the strong surface dependence of the spin splitting. Our findings establish a microscopic mechanism for understanding altermagnetism in reduced dimensions and provide a general principle for engineering spin splitting in low-dimensional altermagnetic materials.
\end{abstract}

\maketitle

Altermagnetism is a recently recognized magnetic phase characterized by momentum-dependent spin splitting despite a vanishing net magnetization \cite{Smejkal-2022-PRX-Emerging, Smejkal-2022-PRX-Beyond, YaoYugui-2025-PRL-Theory, Song-2026-PRM-compensation}. By combining key advantages of ferromagnets and antiferromagnets, altermagnets provide a promising platform for next-generation spintronic devices \cite{Smejkal-2022-PRX-Giant, YaoYugui-2024-PRL-Predictable, YaoYugui-2024-AFM-Exploring, Song-2025-NatureRM-newclass}. Although numerous altermagnetic materials have been theoretically predicted \cite{Guo-2023-MTP-largescale, LuZhongYi-2025-NSR-AIaccelerated, Wan-2025-PRL-HighThroughput, Sufyan-2026-PRM-Highthroughput, Cui-2026-arXiv-DataDriven, Guo-2026-PRB-monolayers}, experimentally realized and well-characterized candidates remain limited \cite{Jiang-2025-NP-KV2Se2O, Zhang-2025-NP-RbVTeO, Takagi-2025-NM-FeS, Bai-2022-PRL-RuO2, Karube-2022-PRL-RuO2, SongCheng-2024-SA-Mn5Si3}. Among them, chromium antimonide (CrSb) has attracted particular attention owing to its giant spin splitting of approximately 1 eV and high Néel temperature ($T_\mathrm{N}\approx700$~K) \cite{Ding-2024-PRL-Large, Snow-1952-PR-Neutron, Kravchuk-2025-PRB-CrSb, Takei-1963-PR-Magnetic}. These outstanding properties make it an ideal system for investigating the interplay between altermagnetism and other factors such as reduced dimensionality and surface orientation.

The realization of altermagnetic spintronic devices relies on the synthesis and understanding of high-quality thin films \cite{Soumyanarayanan-2016-Nature-Emergent}. Recent studies have focused on the electronic structures of CrSb thin films with different surface orientations. Using angular-resolved photoelectron spectroscopy (ARPES) measurment, Reimers \textit{et al.} observed momentum-dependent spin splitting of approximately 0.6 eV along the $P$-$Q$ path in CrSb (10$\bar{1}$0) thin films with a thickness of 30 nm \cite{Reimers-2024-NC-Direct}, while subsequent measurements revealed an even larger splitting of up to 0.91 eV along $\bar{P}$-$\bar{D}$ directions \cite{Liao2025-CPL}. For CrSb (0001) films, Santhosh \textit{et al.} reported bulk-like spin splitting of 0.7 eV with $g$-wave symmetry persisting down to a thickness of 10 nm \cite{Santhosh2025-AM}. Despite these findings, whether altermagnetic spin splitting survives in the thin-film limit remains under debate. For example, Lin \textit{et al.} predicted that a small epitaxial strain can drive CrSb (0001) films into a G-type antiferromagnetic phase with spin-degenerate bands \cite{Lin2026-AM}. More recently, Zhou \textit{et al.} demonstrated symmetry-controlled manipulation of the altermagnetic order in CrSb thin films by engineering the Dzyaloshinskii--Moriya interaction without inducing a magnetic phase transition \cite{Zhou2025-Nature}. These studies highlight the remarkable tunability of CrSb thin films, while the microscopic origin of the effects of reduced dimensionality and surface orientation on altermagnetic spin splitting remains poorly understood.

Meanwhile, theoretical studies have also explored the influence of surface orientation on the electronic structure of CrSb by constructing slab models with different crystallographic terminations \cite{Marfoua2026-PRB,Sorn-2026-arXiv-Projected}. These advances indicate that the altermagnetic spin splitting is highly sensitive to surface-induced symmetry breaking, demonstrating that reduced dimensionality modifies the spin texture. The interplay between the bulk altermagnetic order parameter and the symmetry of the surface determines the character of the spin splitting \cite{Sorn-2026-arXiv-Projected}. For example, surfaces parallel to the bulk nodal planes like (0001) and (10$\bar{1}$0) exhibit spin-degenerate bands, whereas the (2$\bar{1}\bar{1}$0) surface preserves the characteristic $d$-wave spin-splitting symmetry. Other surface orientations display partially compensated spin splitting owing to their distinct crystalline symmetries \cite{Sorn-2026-arXiv-Projected}. Although these surface-dependent spin-splitting phenomena have been reported, the microscopic mechanism responsible for their emergence remains unclear.

In this letter, we systematically investigate the electronic structures of bulk CrSb and ultrathin slabs with various surface orientations using first-principles calculations. We demonstrate that altermagnetic spin splitting is highly sensitive to surface orientation. Our results show that the (2$\bar{1}\bar{1}$0) slab retains a pronounced spin splitting, whereas the (0001) and (10$\bar{1}$0) slabs exhibit nearly spin-degenerate bands. By analyzing the microscopic hopping parameters, we reveal that dimensional reduction fundamentally changes the microscopic origin of altermagnetic spin splitting by making coplanar long-range Cr–Sb interactions indispensable in thin films. Our findings uncover the microscopic origin of surface-dependent spin splitting in altermagnetic CrSb thin films and establish an important design principle for realizing high-performance altermagnetic spintronic devices through surface engineering.

\begin{figure}
\centering
\includegraphics[width=0.46\textwidth]{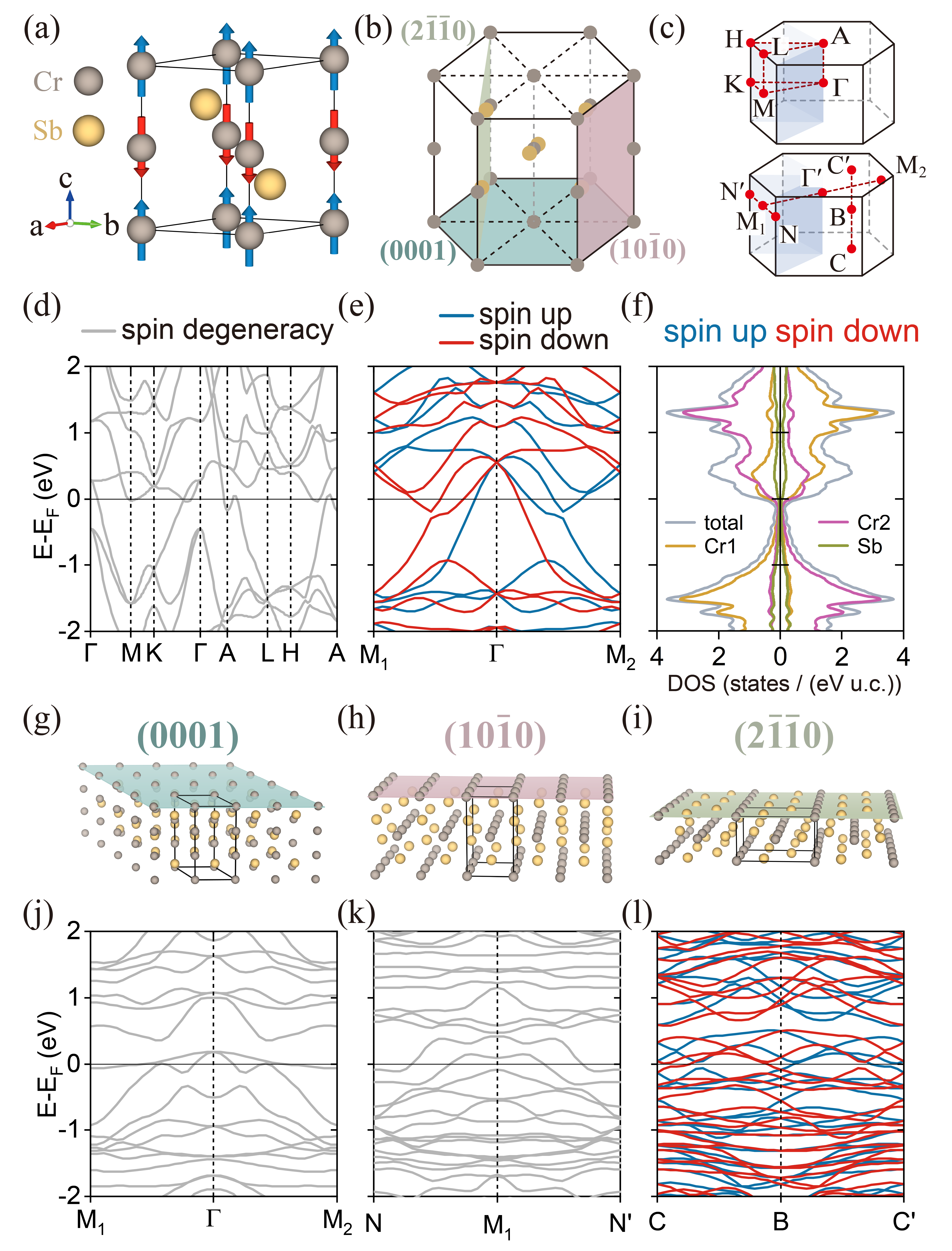}
\caption{(a) Crystal structure and magnetic configuration of bulk CrSb. Blue and red arrows denote the spin-up and spin-down states, respectively. (b) Schematic illustration of the representative crystallographic planes, including (0001), (10$\bar{1}$0) and (2$\bar{1}\bar{1}$0). (c) Brillouin zone of bulk CrSb with the high- and low-symmetry $k$-point paths. (d),(e) Electronic structures of bulk CrSb along the high- and low-symmetry paths. (f) Projected density of states (PDOS) of bulk CrSb. (g)–(i) Atomic structures of CrSb slabs with the (0001), (10$\bar{1}$0), and (2$\bar{1}\bar{1}$0) surface orientations, respectively. (j)–(l) Electronic structures of the corresponding CrSb slabs calculated along the low-symmetry $k$-point paths, respectively.}
\label{fig1}
\end{figure}

Bulk CrSb crystallizes in the hexagonal NiAs-type structure with space group $P6_3/mmc$ (No.~194), as illustrated in Fig.~\ref{fig1}(a) \cite{Zeng2024}. The spin-up and spin-down sublattices are related by a $\pi/3$ rotation combined with a half-lattice translation ($t_{1/2}$) along the [0001] direction or, equivalently, by mirror reflection with respect to the (0001) plane. The electronic structures were calculated using the Vienna \textit{Ab-initio} Simulation Package (VASP) \cite{Kresse-1996-PRB-PW, Blochl-1994-PRB-PAW, Kresse-1999-PRB-PAW}. The calculated band structures along the high- and low-symmetry $k$ paths of the Brillouin zone [Fig.~\ref{fig1}(c)] are presented in Figs.~\ref{fig1}(d) and \ref{fig1}(e). As expected for an altermagnet, pronounced momentum-dependent spin splitting is observed along the low-symmetry $k$ path on the $k_z=0.25\times c^*$ plane, whereas spin degeneracy is preserved along the nodal planes. To investigate the influence of reduced dimensionality and surface orientation, we further construct CrSb slab models with (0001), (10$\bar{1}$0), and (2$\bar{1}\bar{1}$0) surface orientations, separated by a vacuum region of 15~\AA, as shown in Figs.~\ref{fig1}(g)–\ref{fig1}(i). The corresponding crystallographic planes are illustrated in Fig.~\ref{fig1}(b).

A striking contrast emerges in the electronic structures of the CrSb slabs with different surface orientations. As shown in Figs.~\ref{fig1}(j) and \ref{fig1}(k), the momentum-dependent spin splitting disappears in the (0001) and (10$\bar{1}$0) slabs, resulting in nearly spin-degenerate bands. In contrast, the (2$\bar{1}\bar{1}$0) slab retains the characteristic altermagnetic spin splitting in Fig.~\ref{fig1}(l). Since all three slab models preserve the altermagnetic spin-group symmetry \cite{amcheck, FindSpinGroup}, the markedly different electronic structures cannot be attributed solely to symmetry. Furthermore, the calculated magnetic moments of the Cr atoms vary slightly among the different slabs and remain close to 3 $\mu_{\rm{B}}$, indicating that the suppression of spin splitting is not associated with a significant change in magnetic moments. Similar surface-dependent behavior has recently been reported in studies of CrSb surface states \cite{Sorn-2026-arXiv-Projected}. These observations therefore suggest that additional factors beyond magnetic symmetry govern the emergence of spin splitting in CrSb slabs.

The microscopic origin of altermagnetic spin splitting is closely related to the electronic interactions between magnetic and nonmagnetic atoms under the constraints of the crystal symmetry and magnetic order. Consequently, the electronic bonding and orbital hybridization play a crucial role in determining the spin splitting. As shown by the projected density of states in Fig.~\ref{fig1}(f), the Cr-$3d$ and Sb-$5p$ orbitals exhibit similar spectral features in the vicinity of the Fermi level, indicating pronounced Cr--Sb hybridization. The markedly different spin-splitting behaviors observed in bulk CrSb and slabs with different surface orientations therefore suggest that the underlying Cr--Sb interactions are strongly modified by reduced dimensionality. To elucidate this effect, we systematically investigate the evolution of the electronic structures by varying the vacuum spacing between neighboring slabs, thereby continuously tuning the coupling from the three-dimensional bulk limit toward the two-dimensional thin-film limit.

\begin{figure}
\centering
\includegraphics[width=0.46\textwidth]{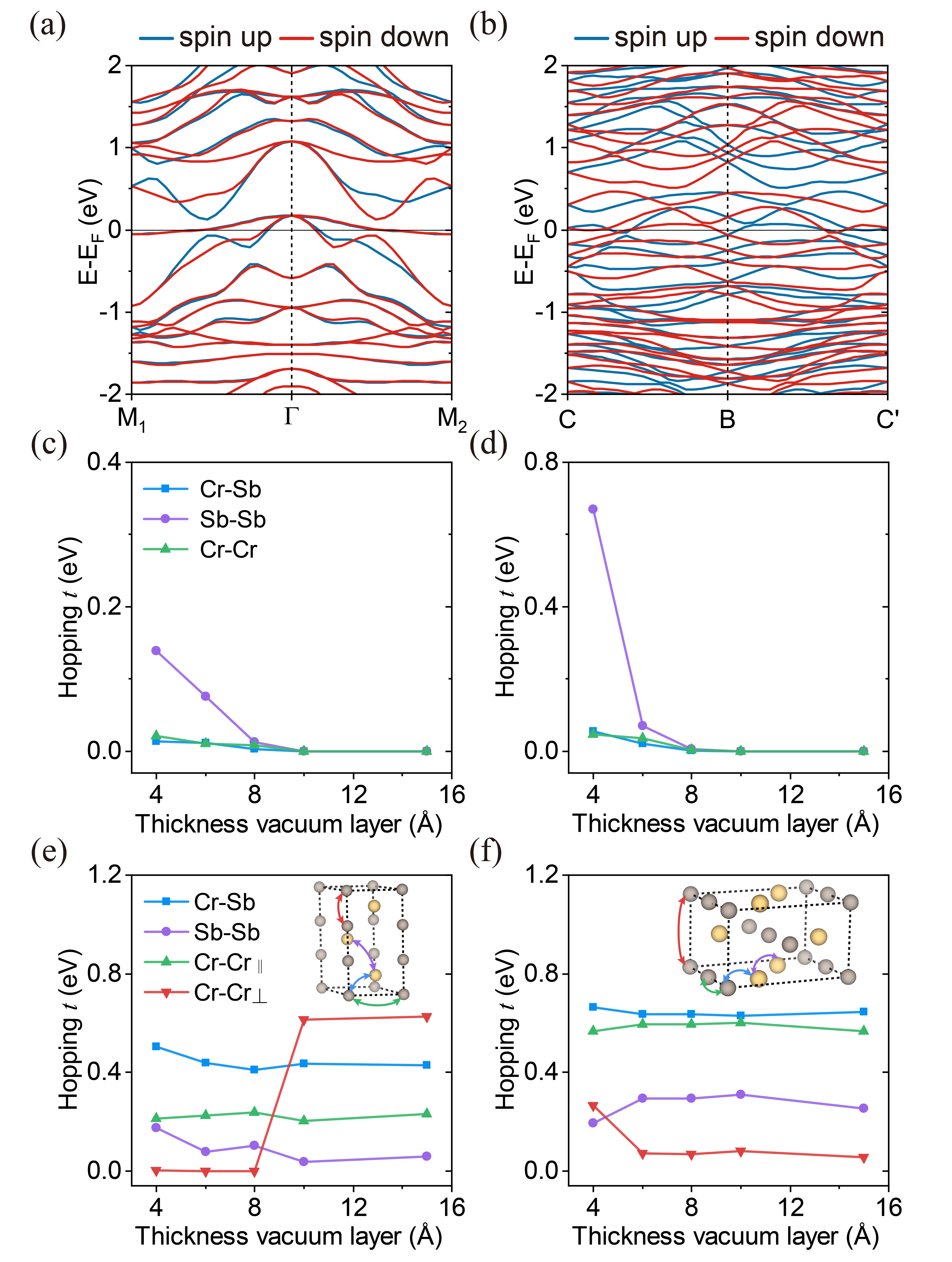}
\caption{(a),(b) Electronic structures of the CrSb (0001) and (2$\bar{1}\bar{1}$0) slab calculated with a vacuum spacing of 4~\AA\, respectively. (c),(d) Schematic illustration of the inter-slab Cr–Sb, Sb–Sb, and Cr–Cr hopping interactions across the vacuum region for the (0001) and (2$\bar{1}\bar{1}$0) slabs, respectively. (e),(f) Schematic illustration of the Cr--Sb, Sb--Sb, and Cr--Cr intra-slab hopping interactions within the (0001) and (2$\bar{1}\bar{1}$0) slabs.}
\label{fig2}
\end{figure}

We focus on the (0001) and (2$\bar{1}\bar{1}$0) surface orientations for they exhibit distinctly different spin-splitting behaviors in Fig.~\ref{fig1}. The different evolution of the electronic structures is illustrated in Fig.~\ref{fig2}, as the vacuum spacing is varied. For the (0001) slab, reducing the vacuum spacing to 4~\AA\ restores momentum-dependent spin splitting in Fig.~\ref{fig2}(a), although its magnitude remains substantially smaller than that of bulk CrSb. Notably, the overall band dispersion changes only slightly with the vacuum spacing. In contrast, the (2$\bar{1}\bar{1}$0) slab retains pronounced spin splitting for both 4~\AA\ and 15~\AA\ vacuum spacings, while the electronic band structure evolves with increasing vacuum spacing in Fig.~\ref{fig2}(b). The different evolution of the electronic structures for the two surfaces indicates that reduced dimensionality modifies the underlying electronic interactions in a highly anisotropic manner. These results suggest that the microscopic interactions responsible for altermagnetic spin splitting are largely suppressed in the (0001) slab but remain robust in the (2$\bar{1}\bar{1}$0) slab.

To elucidate the microscopic origin of the evolution of spin splitting with vacuum spacing, we constructed Wannier tight-binding Hamiltonians for the CrSb slabs using Cr-$3d$ and Sb-$5p$ orbitals within the WANNIER90 package \cite{Pizzi-2020-IOP-Wannier90}. For clarity, Fig.~\ref{fig2} displays only the hopping parameters with the largest magnitudes. As summarized in Figs.~\ref{fig2}(c) and \ref{fig2}(d), these dominant inter-slab hopping amplitudes decrease rapidly with increasing vacuum spacing and become nearly negligible beyond approximately 8~\AA. This evolution reflects the progressive suppression of inter-slab electronic coupling, driving the periodic slab model toward the two-dimensional limit.

In contrast, the evolution of intra-slab hopping parameters exhibits distinct behaviors for the (0001) and (2$\bar{1}\bar{1}$0) slabs. As shown in Fig.~\ref{fig2}(e), most intra-slab hopping amplitudes in the (0001) slab remain nearly unchanged as the vacuum spacing increases. However, the Cr--Cr$_{\perp}$ hopping, corresponding to the direction perpendicular to the surface, increases significantly when the vacuum spacing reaches 8~\AA, where the inter-slab hopping becomes negligible. This anomalous enhancement suggests a redistribution of the dominant electronic coupling pathways caused by the removal of surface-to-surface interactions. In other words, the elimination of inter-slab coupling modifies the remaining intra-slab electronic network and strengthens the Cr--Cr$_{\perp}$ channel. In contrast, as shown in Fig.~\ref{fig2}(f), the intra-slab hopping parameters in the (2$\bar{1}\bar{1}$0) slab exhibit only weak variations with increasing vacuum spacing. In particular, the Cr--Cr$_{\perp}$ hopping decreases slightly, indicating that the essential electronic coupling pathways are preserved in this surface orientation. These contrasting behaviors suggest that the distinct intra-slab hopping networks in different surface orientations play a crucial role in determining the presence or absence of altermagnetic spin splitting.

\begin{figure}
\centering
\includegraphics[width=0.47\textwidth]{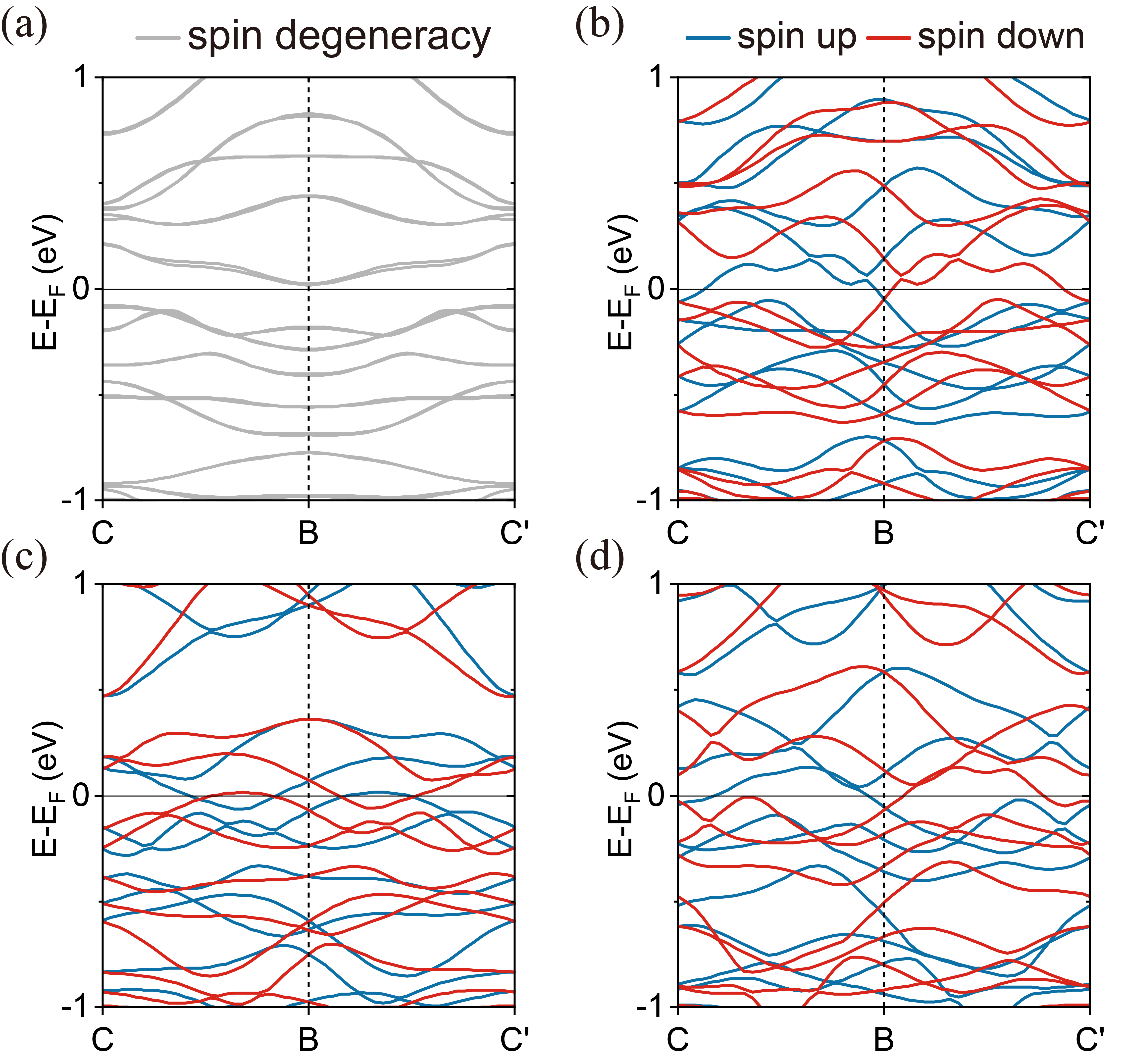}
\caption{Electronic structures obtained from the Wannier tight-binding Hamiltonian under different hopping configurations while preserving the altermagnetic symmetry. (a) The truncated Wannier Hamiltonian by retaining only the intra-unit-cell hopping terms. (b) The truncated Wannier Hamiltonian including Sb--Sb hopping parameters connecting neighboring unit cells. (c) The truncated Wannier Hamiltonian including Cr--Sb hopping parameters connecting neighboring unit cells. (d) The truncated Wannier Hamiltonian including Cr--Sb and Sb--Sb hopping parameters connecting neighboring unit cells.}
\label{fig3}
\end{figure}

Previous studies have suggested that third-nearest-neighbor Cr--Cr hopping is responsible for the altermagnetic band splitting in bulk CrSb \cite{Yang-2025-NC-mapping}. In contrast, the study of the isostructural compound NiS has demonstrated that interactions involving nonmagnetic atoms play a dominant role in generating spin splitting \cite{Mandal-2025-PRB-NiS}. Since these studies did not systematically analyze the contributions of different hopping channels, the microscopic interactions responsible for altermagnetic spin splitting are unclear. When the dimensionality is reduced from bulk material to slabs, the issue becomes more complex due to the surface modifying the hopping network. Here, we focus on the (2$\bar{1}\bar{1}$0) slab to identify the dominant hopping pathways responsible for the observed spin splitting. As a starting point, we construct a truncated Wannier Hamiltonian by retaining only the intra-unit-cell hopping terms allowed by the altermagnetic symmetry. As shown in Fig.~\ref{fig3}(a), despite preserving the dominant short-range hopping interactions within the unit cell, the resulting electronic structures exhibit nearly spin-degenerate bands. Notably, although the third-nearest-neighbor Cr--Cr hopping is included in this model \cite{Yang-2025-NC-mapping}, no significant spin splitting emerges in the (2$\bar{1}\bar{1}$0) slab. This result demonstrates that the microscopic origin of spin splitting in the thin-film limit differs fundamentally from that in bulk CrSb.

We next selectively restore the hopping terms connecting neighboring unit cells to identify the microscopic origin of the spin splitting. As shown in Fig.~\ref{fig3}(b), including the Sb--Sb hopping between neighboring unit cells is sufficient to induce momentum-dependent spin splitting, indicating that long-range inter-unit-cell hopping is indispensable for the emergence of altermagnetic spin splitting in the low-dimensional (2$\bar{1}\bar{1}$0) slab. However, the resulting band dispersion still deviates significantly from the fully reconstructed electronic structure. In contrast, when only the Cr--Sb hopping between neighboring unit cells is retained in Fig.~\ref{fig3}(c), both the overall band dispersion and the spin splitting closely reproduce the first-principles results shown in Fig.~\ref{fig1}(l). This demonstrates that the long-range Cr--Sb hopping provides the dominant contribution to the altermagnetic spin splitting in the (2$\bar{1}\bar{1}$0) slab. Finally, including both the Cr--Sb and Sb--Sb hopping terms between neighboring unit cells nearly reproduces both the electronic structure and spin splitting, as shown in Fig.~\ref {fig3}(d).

\begin{figure}
\centering
\includegraphics[width=0.47\textwidth]{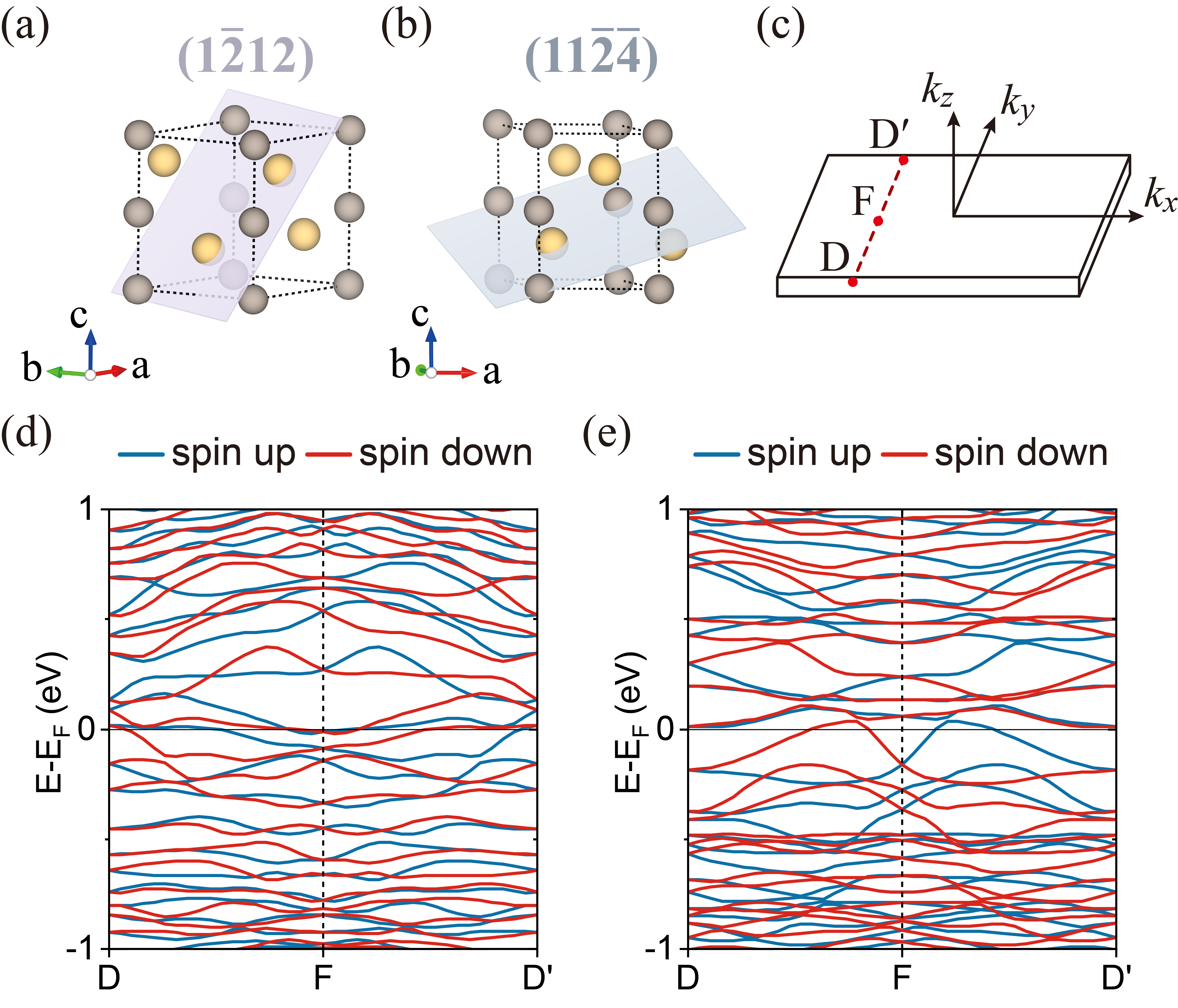}
\caption{(a),(b) Schematic illustrations of the CrSb with the (1$\bar{2}$12) and (11$\bar{2}$$\bar{4}$) surface orientations, respectively. (c) Brillouin zone of the (1$\bar{2}$12) and (11$\bar{2}$$\bar{4}$) slabs, with the low-symmetry $k$-path indicated by the red dashed line. (d),(e) Electronic structures of the CrSb (1$\bar{2}$12) and (11$\bar{2}$$\bar{4}$) slabs.}
\label{fig4}
\end{figure}

The hopping analysis of the (2$\bar{1}\bar{1}$0) slab identifies the long-range inter-unit-cell Cr--Sb hopping as the dominant interaction responsible for the altermagnetic spin splitting, with the Sb--Sb hopping providing an additional contribution. This microscopic picture can be understood directly from the crystal structure shown in Fig.~\ref{fig1}(i). In the (2$\bar{1}\bar{1}$0) slab, Cr and Sb atoms form well-defined coplanar hopping pathways within the same (2$\bar{1}\bar{1}$0) plane, allowing the long-range Cr--Sb interactions to be preserved even in the thin-film limit. In contrast, the Cr and Sb atoms in the (0001) slab are located at different depths relative to the surface, making the corresponding hopping pathways much more susceptible to dimensional reduction. This interpretation is consistent with the pronounced enhancement of the Cr--Cr$_{\perp}$ hopping shown in Fig.~\ref{fig2}(e), which indicates a substantial redistribution of the intra-slab hopping after the inter-slab coupling is suppressed. It also naturally explains the recovery of spin splitting when the vacuum spacing is reduced in the (0001) slab. Notably, the inter-slab hopping amplitudes become nearly negligible for vacuum spacings of approximately 8~\AA\ in Fig.~\ref{fig2}(c), just comparable to the spatial extent of the inter-unit-cell Cr--Sb hopping pathway. These results demonstrate that, unlike bulk CrSb, altermagnetic spin splitting in thin films requires long-range inter-unit-cell hopping, with the coplanar Cr--Sb hopping providing the dominant microscopic mechanism.

To further validate the role of Cr--Sb hopping in generating altermagnetic spin splitting, we additionally investigate the electronic structures of the CrSb (1$\bar{2}$12) and (11$\bar{2}$$\bar{4}$) slabs. The corresponding surface orientations are illustrated in Figs.~\ref{fig4}(a) and \ref{fig4}(b). For the (11$\bar{2}$$\bar{4}$) slab, the Cr and Sb atoms can form coplanar hopping pathways parallel to the surface, allowing the long-range Cr--Sb interactions to be largely preserved. Consequently, pronounced spin splitting is retained, as shown in Fig.~\ref{fig4}(e), although its magnitude is slightly smaller than that of the (2$\bar{1}\bar{1}$0) slab because of the larger Cr--Sb separation. In contrast, the Cr and Sb atoms in the (1$\bar{2}$12) slab are not coplanar. Nevertheless, a finite spin splitting remains visible in Fig.~\ref{fig4}(d), indicating that the Cr--Sb hopping is not completely suppressed. This can be attributed to the relatively small separation between the Cr and Sb atomic planes, which allows appreciable interlayer Cr--Sb hopping to persist despite the presence of the surface. The weaker spin splitting in the (1$\bar{2}$12) slab compared with the (11$\bar{2}$$\bar{4}$) slab further supports this interpretation. Taken together, these results consistently demonstrate that long-range Cr--Sb hopping is the key microscopic interaction governing altermagnetic spin splitting in CrSb thin films across different surface orientations.

In summary, we have systematically investigated the electronic structures of bulk CrSb and slabs with different surface orientations. We demonstrate that altermagnetic spin splitting exhibits a pronounced dependence on surface orientation. The (2$\bar{1}\bar{1}$0) and (11$\bar{2}$$\bar{4}$) slabs retain robust spin splitting, whereas the (0001) and (10$\bar{1}$0) slabs exhibit nearly spin-degenerate bands, with only weak splitting remaining in the (1$\bar{2}$12) slab. By selectively analyzing the microscopic hopping channels, we reveal that dimensional reduction fundamentally changes the microscopic origin of altermagnetic spin splitting. The thin films require long-range inter-unit-cell Cr--Sb hopping, while the Sb--Sb hopping provides an additional contribution. The preservation or suppression of these hopping pathways is determined by the surface geometry, giving rise to the pronounced surface dependence of the spin splitting. Our work establishes a microscopic mechanism for understanding altermagnetic spin splitting in reduced dimensions and provides a general strategy for engineering altermagnetic electronic structures through surface and interface design.

This work is supported by the National Key Research and Development Program of China under Award (Grant No. 2021YFA1201800), the National Natural Science Foundation of China (Grants No. 52371174 and No. 12204033), the Science Challenge Project (Grant No. TZ2025009), and the State Key Lab of Advanced Metals and Materials (Grant No. 2025Z-Z22). Numerical computations were also performed on Hefei advanced computing center.

\nocite{*}
\bibliography{CrSbref}

\end{document}